# Identifying Benefits and Risks Associated with Utilizing Cloud Computing

Jafar Shayan[1]
Advanced Informatics School, Universiti Teknologi Malaysia
sjafar3@live.utm.my

Ahmad Azarnik[2]
Advanced Informatics School, Universiti Teknologi Malaysia
aahmad8@live.utm.my

Suriayati Chuprat[3]
Advanced Informatics School, Universiti Teknologi Malaysia
suria@ic.utm.my

Sasan Karamizadeh[4]
Advanced Informatics School, Universiti Teknologi Malaysia,
ksasan2@live.utm.my

Mojtaba Alizadeh[5]
Malaysia-Japan International Institute of Technology, Universiti Teknologi Malaysia
amojtaba2@live.utm.my

*Abstract*— Cloud computing is an emerging computing model where IT and computing operations are delivered as services in highly scalable and cost effective manner. Recently, embarking this new model in business has become popular. Companies in diverse sectors intend to leverage cloud computing architecture, platforms and applications in order to gain higher competitive advantages. Likewise other models, cloud computing brought advantages to attract business but meanwhile fostering cloud has led to some risks, which can cause major impacts if business does not plan for mitigation. This paper surveys the advantages of cloud computing and in contrast the risks associated using them. Finally we conclude that a well-defined risk management program that focused on cloud computing is an essential part of gaining value from benefits of cloud computing.

## I. INTRODUCTION

Cloud computing has changed the information technology (IT) services in the several ways: development, deployment, scalability, update, maintenance, and payment. The expense of computing has increased in an organization due to increase of complexity of management, infrastructure of information architecture, and distributed data and software[1]. Cloud computing is able different services for information technology and reduce the cost for small and medium companies that are unable to deploy and use many cutting-edge IT services[2]. Gartner research expects that investment in cloud computing will be the $150 billion. One survey on six datacentre showed that servers used only 10-30% of their computing power, while desktop computers utilized less than 5%[3]. Organizations and companies spend about two-third of IT staffing budget on support and maintenance activities[4], which seems unnecessary in the age of globalization.

Cloud computing represents two trends: IT efficiency and business agility. IT efficiency enables vendors and users to employ the power of modern computer efficiently through large scalable hardware and software. It also carries the concept of green computing; applications can be run in computers with different geographical areas by cheap electricity .Business agility makes development, parallel processing, business analysis, and mobile interactive application to accessible for users in real time[5]. It reduce the cost of enterprise IT setups by means of computational tools that can be deployed and scaled rapidly.

According to National Institute of Standards and Technology (NIST): "Cloud computing is defined as a model for enabling convenient, on-demand network access to a shared pool of  configurable computing resources (e.g., networks, servers, storage, applications, and services) that can be  rapidly provisioned and released with minimal management effort or service provider interaction"[6].

Cloud computing can be declared as a model of delivering IT services to customers on demand through internet or network (in private or hybrid deployment model) in a self-service manner. Basic principal behind this new model is sharing computing resources and delivering scalable services regardless of local as this model is based on virtualization. This model offers varied pricing models meeting diverse business requirements but most important one is pay as you go for resources which have been utilized and no need to invest high amount of capital as upfront cost.

Virtual technology manages the cloud systems which mean that on the same hardware different users can access different operation systems and services. Users are concerned about the security of cloud systems, such as data are separated securely, accessible on-demand, who are responsible for data lost and etc[3, 4]. According to IDCI, 74% of IT executives and CIOs mentioned the security as a burrier to move into cloud service model. As the global market and businesses move to cloud systems privacy and risks must be addressed [7]. In section 2 we bring an explanation of cloud computing; section 3 includes benefits of cloud computing; and in section 4, we categorize the risks-related cloud computing regarding to tangible and intangible.

## II. WHAT IS CLOUD COMPUTING?

For several years, CIOs were freely working and their focus was IT effectiveness. Despite CFOs unhappiness of spending much capital on IT, they were spreading IT infrastructures and try to adjust all these fees by saying it can provide competitive advantages but after economic recession, CIOs felt heavy pressure on them to justify expenditures and business advantages. Growing IT infrastructure needs more support human resource and even using different technology and tools need different expertise, which costs more and makes it complicated. By decreasing IT budget, it was obvious that cost effective solutions would become hot topics[8].

The philosophy behind cloud computing is very simple. Get rid of any IT burden. Utility computing is another definition which makes it clear that it would become 5th utility[9].

By declaring IaaS (Infrastructure as a Service), PaaS (Platform as a Service), SaaS (Software as a Service) new age of computing emerged. Cloud providers like Amazon, Microsoft, Google, etc. delivering computing services as utility.

This new service model showed that can bring some benefits like cost reduction by declaring Pay as you Go which is base model for utility model, improve provisioning and access to resources and availability[10].

But going to clouds without considering all potential risks is not wise decision. Although top managers want to reduce costs but they don't like to lost their business by underestimating security risks.

## III. CLOUD COMPUTING BENEFITS

Since computing emerged and found its significant role in business competitive, the penetration of computing within organizations increased dramatically. By increasing the usage of computing as an important initiative, nowadays it can see as a must to use. Meanwhile, complexity if the computing architecture and information distributed in diverse locations as well as different business units and functionalities, brought new challenges and costs for organizations. Considering these challenges, Cloud computing showed up as a solution to deliver IT functionalities among organizations as services with the minimum upfront cost and complexity.

For several years, CIOs were freely working and their focus was IT effectiveness. Despite CFOs unhappiness of spending much capital on IT, they were spreading IT infrastructures and try to adjust all these fees by saying it can provide competitive advantages but after economic recession, CIOs felt heavy pressure on them to justify expenditures and business advantages. Growing IT infrastructure needs more support human resource and even using different technology and tools need different expertise, which costs more and makes it complicated. By decreasing IT budget, it was obvious that cost effective solutions would become hot topics[2].

The philosophy behind cloud computing is very simple. Benefit from IT as it can play enabler role inside organization and not worry about it because provider will care about it. Utility computing is another definition, which makes it clear that it would become 5th utility[9].

### A. COST

In a "Cloud Migration: A Case Study of Migrating an Enterprise IT System to IaaS", Khajeh-Hosseini et al. [11] talked about the third party cloud infrastructure and highlighted the benefits of embarking third party cloud computing.

They showed that using third party cloud computing offers different advantages in terms of income and outgoing management improvement for both side of customers and finance staff. Reducing the upfront cost which is introduced by cloud computing and offering the new price model for IT related expenditures, which is monthly billing model, helps easing the cash flow. Moreover, outsourcing computing needs to third party cloud infrastructure reduced the electricity expenditures. They extracted these benefits by comparing the outsourcing IT services to third party cloud provider rather than running those services on in-house data centers. By transferring the responsibility of IT services to third party provider, company ran away of buying and upgrading servers and required hardware and software. In this scenario energy consumption of IT related services dramatically decreased as they are not running in-house anymore. Interestingly, outsourcing those services reduces the administrative burden by putting it on third party's shoulder and this makes finance department completely satisfied. Offering new and different pricing models by third party cloud solutions help different stakeholders to manage incomes more efficient including customers, sales and marketing staff [11].

Economics and simplification which indirectly will result in cost reduction are introduced as those can be listed in main drivers of cloud computing [12]. Dorey and Leite[10] mentioned that cost reduction one of the items that drive the IT environment to go for cloud computing. In providing a future perspective of cloud computing, Lillard et al [13] announced that supplying on demand computing power in a very low-cost fashion was the main driver of cloud computing emergence. Furthermore, Marston et al [14] highlighted the lower cost of entry for small business as one of key advantages of cloud computing not only for SMEs but for third world countries as well. Cloud computing represents a huge opportunity to many third-world countries that have been so far left behind in the IT revolution by reducing IT barriers for them including the cost [14]. Also they pointed out that cloud computing does not need heavy upfront capital investment as another benefits offering by cloud computing.

### B. AGILITY

In highly competitive business environment of nowadays, obtaining the competitive advantages are vital necessary. Among those initiatives, a key competitive differentiator is the ability to respond to customer needs, which is fast changing. Cloud provides the ability to adjust processes, products and services in timely manner. Therefore, business can meet the market change requirements by increasing the agility[15].

Agility can be introduced by focusing in business processes and improve them. Cloud computing offers agility by offloading three kinds of low-level administration to cloud providers: system infrastructure, backup policy, and single application. Maintaining system infrastructure includes hardware maintenance, adding new machines and upgrading current machines, spare parts management and infrastructure software can be handled by cloud provider. Second, cloud provider is responsible for managing backup related activities according the backup policy defined by organization. Lastly, having a single application for serving to all authorized users, make the software management and all related activities including application support, upgrade issues and user management easier and cost effective [16].

Cloud computing can increase the possibility of going for innovation through lowering IT barriers. Many successful startups such as Facebook and YouTube and some more focused applications [16]. Offloading the irrelevant IT concerns which are not adjusted to business processes, increase opportunity of business process and operations improvement by innovative solutions.

### C. GREEN COMPUTING

Computing power including server powers, cooling and overhead power consumption is expensive. It can be worse when power management is not well done or is not efficient. Cloud vendors are managing these much better comparing typical on promise data centers and legacy data centers based on effective management of voltage conversion, spending less on cooling by locating data centers in cooler places and having better cooling facilities and also lower electricity rates (cloud vendors tend to cluster near hydropower)[16]. They often locate where natural facilities help them do cooling easier and with less energy consumption.

Conventional data centers are suffering of low resource utilization which highest estimates are 15-20% while there are lower estimates. In cloud computing data centers this situation is better by having multi tenants and sharing resources for many customers upper the resource utilization to 40% by load sharing over time zones, mature virtualization leveraging and fostering more diverse user bases. One virtual server can handle work of at least 2.5 typically utilized servers [16]. Consequently, higher utilization means less power wasting and utilizing valuable power efficiently, which helps the environment to keep safe.

An IB energy assessment was done in 2008 within the whole world. Interestingly results showed indirect purposes was the first energy consumption in data centers with more than 60-70% of whole energy usage. Direct energy usage is about 30-40 % while others go for cooling and lightening the facilities [17]. Public clouds are doing the most efficient energy use by locating their data centers in optimal places in terms of bandwidth, energy, abundant water for cooling, proximity to market and cheaper human resources. Meanwhile they are researching on approaching creative ideas to decrease resource usage not only electricity usage, but also water recycling as well as equipment recycling [18].

### D. EMERGING APPLICATIONS

Knowing highlighted benefits earlier, researchers and practioners address more advantages for using cloud computing such as scalability of services [15, 16, 18], different billing types [15].

New classes of application emerged by leveraging the opportunity that cloud computing provide. These applications were not possible before that but huge processing power and cheap price rate of services allow location-environment and context aware applications exist. These applications need respond in real time to information provided by human users, nonhuman sensors such as humidity and stress sensors within a shipping container [16]. Another application type is parallel batch processing which is allows user to analyze huge amount of data in a relatively short period by leveraging huge processing power provided by cloud computing infrastructure [16].

Business analytics is very attractive for businesses and bring them competitive advantage of running smart business by processing vast amount of data to understand and predict the customer behavior, market analysis, buying habits and so on is another benefit of saving data in cloud and make it easier and applicable [15, 16].

During the time, IT and computing resources became complex and complicated and benefiting them, needs deep technical expertise, knowledge and skills which consist of higher expenses. Cloud computing is offloading unnecessary sophisticated IT related tasks and management, hides the complexities. Therefore, end users can produce sophisticated products or services without high level IT expertise requirement [15].

Cloud computing offers a fair competition environment and lowering IT barriers. Large companies used to be winner in competitions against smaller firms benefiting higher resource in terms of human capital, software and hardware resources to support new marketing and strategic initiatives. However, cloud computing lower the barriers by decreasing upfront costs and investment costs as well as new pricing models which fit small companies and startups [18]. These facts are not only relevant to small middle enterprises (SMEs) but also developing countries can benefit cloud services and cover their limitations regarding with limited available capital, resources and skilled humans.

### IV. ASSOCIATED RISKS OF CLOUD COMPUTING

Although cloud computing has many benefits and opportunities, there are several risks that should be consider before migrating to it. Knowing the risks and well planned management of them can assure the Confidentiality, Integrity and Availability of data (CIA) [19]. Without determining risks for both provider and users, if something is done wrongly, it will lead to bankruptcy and losing the market [20].

As Figure 2 shows, we separate risks to two categories, tangible and intangible. *Tangible risks* are those that users could easily understand them and decide to select the appropriate ones. The risks that are transparent to users and only provider know them; these risks are called *intangible risks* [21].Figure-2

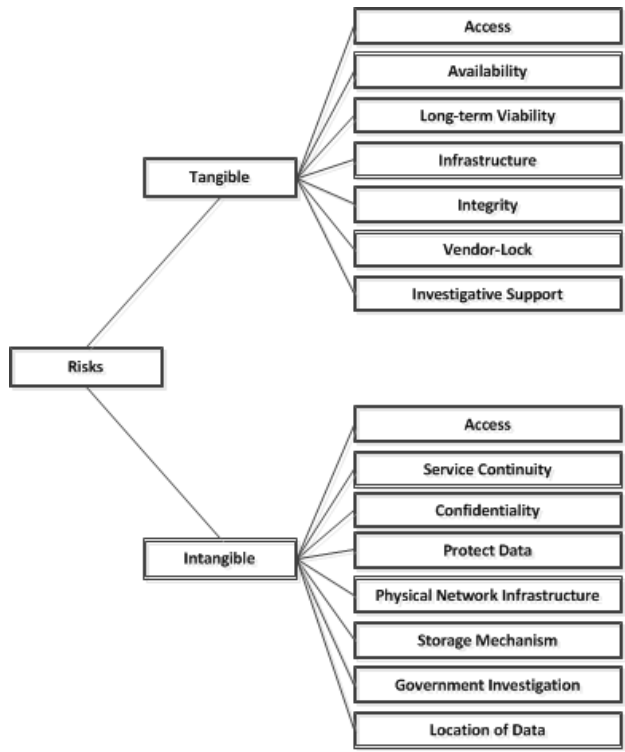

Figure 2 Tangible and intangible risks

### A. TANGIBLE RISKS

In cloud computing, it is possible that the storage is in one place and the process is done in another server by different geographical position. However, it does not mean the clients are not familiar to everything. The *Tangible Risk* is a thing that customers could easily understand and touch. In this paper, the tangible risks contain access, availability, Long-term viability, Infrastructure, Integrity, Vendor lock-in and Investigative support.

*1. Access*

Data should be secured and only authorized users can access data. The cloud computing needs to be equipped with good and functional mechanism for identification, authentication and authorization. Identification recognizes the individual users, authentication proves that a user is true when he or she claims, and authorization assures that user is authorized properly and is able to access, update or delete information [22]. In cloud computing environment that is based upon multi-tenant service providing, it is critical to ensure right persons access to their data in easy and quick regardless of geographical places and there should be a mechanism to protect their data against unauthorized data access.

*2. Availability*

In enterprise scale, customers expect to have no interruption in availability, because it directly affects their business. In education, may be some courses are done in the cloud, online teaching, doing assignment and even tests and quiz. It becomes worst when a conference is held, there are lot of articles should be presented and some speakers do it online.

The reasons could be an interruption in electricity, a failure of hardware, or a natural disaster, flood, earthquake, Tsunami to name a few. The cloud computing is a good aim for hackers who can compromise the system, steal data and disable the way to access services.

*3. Long-term viability*

The provider may become bankrupt and goes out of business or sell to another company. You should ask providers how you can catch your data if it happens. In 2008, the "Linkup" suddenly left the business and shocked its 20000 customers. Officials claimed that 55% of data was safe and the rest was not clear [21].

This risk exist in other IT outsourced projects or services and needs comprehensive research on selecting provider and obtaining detailed service level agreement options addressing this issue and controlling the provider situation as well as having contingency plan.

*4. Infrastructure*

The flexibility and scalability are essential for cloud infrastructure. If the infrastructure is not implemented correctly, the growing of the cloud could cause bad and irreparable effects on the vendor. For instance, upgrading hardware and updating software. In education, it become more important when school or university intend to test a complex examination on the cloud, for this aim, the cloud vendor may have to change some things in the infrastructure, without a flexibility of the system it is not possible [21].

Infrastructure should be isolated from bad effects of tenants on other parts of system. Although base concept of cloud computing is integrating computing resources, but it should be consider that without flexible and scalable infrastructure, user's requirements will not meet and subsequently it will fail.

*5. Integrity*

Integrity means that data remains whole complete and unchanged when data is transferred, processed and stored. The corruption, damage and destruction of data are not acceptable by customer [22].

The providers must ensure that data in their cloud transmitted without any problem and it should be mentioned in SLA. Moreover, in recover data integrity must be assured, when data is corrupted or lost. The vendor may try to recover the 50% of information, and for the rest, it is probable that data is either never recovered or recovered in corrupted files. Customers have to wait for days without any warranty of achieving proper data. The provider by selecting a good and effective integrity approach can reduce cost and time.

*6. Vendor lock-in*

Cloud computing offers convenient services and this point is the reason of cloud popularity. Interestingly, it can bring vendor control. This concern is usual in outsourcing projects when customer highly relies on particular provider or the architect is based on a certain vendor [23]. For avoiding this issue the solution usually comes with vendor independency. Costly switching to another cloud provider makes organizations hesitate about considering this concern. Even though there are many choices but immense switching costs and time consuming migrations activities plus sophisticated technical knowledge need, make customers tie to particular provide [24].

*7. Investigative support*

Forensic investigation gathers the facts and evidence related to the crime. This includes software, hardware, and user or vendor. The essence of cloud computing makes it hard and most times impossible to investigate the fraud and hacking because logging and data for multiple customers may be co-located and also may be spread across an ever-changing set of hosts and data centers [25]. The difficulties include: identification of the source and infected systems, collection and seizure of evidence, preservation, analysis, reconstruction, and reporting.

B. INTANGIBLE RISKS

There are instances that users could not easily understand the problems, and vendors are responsible to provide them. These kinds of risks are mentioned as Intangible Risks which includes *access, Service continuity, Confidentiality, Protect data, Physical network infrastructure, Storage mechanism, Government investigation and Location of data.*

*1. Access*

Here it means that users can access data wherever and whenever they want. The vendors should provide connection in which way that no interruption and disconnection happen because cloud computing requires a reliable internet connection [26]. In fact most cloud providers prepare redundant connection to prevent this problem and it is more likely to see break connection on customer's end.

*2. Service continuity*

A loss of one service could not affect the function of the system and users. Nobody should notice the failure in a service [21]. This includes regular backup, updating anti viruses, inspecting redundant systems and communication lines. Some process of education needs to be done sequentially, for instance, during the enrolment, an interruption in financial system leads to stop the work proceeding.

*3. Confidentiality*

Every moment, thousands of data travel across the cloud system, if they are sent in clear or plain text, hackers can easily sniff information. Nowadays, data encryption is one of the most important concerns; transferring data without any encryption put the customers' data in danger. A lot of mechanism and encryption algorithms exist that can be used to guarantee the secrecy. Most famous algorithms are AES, DES, 3DES and public key [26].

*4. Protect data*

It includes software and hardware used to back up data, or protect them against attacks and malware. For back up the cloud vendors need redundant system and storage in safe place. Protecting the cloud and data requires firewalls, Intrusion detection system (IDS), intrusion prevention system (IPS) and antivirus. Without them, the provider may encounter irreparable damages.

*5. Physical network infrastructure*

It guarantees that data among servers and users travel with suitable and acceptable speed. Although the data links become so fast, many companies lend slower link for reducing the cost for themselves and the best offer for clients. When packets move from source to destination, the speed of data transfer should not affect applications and services.

*6. Storage mechanism*

The provider may use different format and storage technologies. The format must be changeable to another format, it is needed when customer decides to change the vendor or even vendor does not continue its business anymore [21]. Cloud computing vendors need to design a good infrastructure for storage because by doing it the capacity quickly and economically can be expanded. In addition, services could be supplied quickly and more flexible and cost-effective [27].

*7. Government investigation*

The government has been able to demand some details of your online activities from service providers after obtaining search warrant - and not to tell you about it. In the U.S.A the law was passed which allows the investigators to ask about information about users. There have been a lot of requests since the law was passed, and the F.B.I.'s inspection have shown that there can be plenty of overreach perhaps completely inadvertent in requests.

*8. Location of data*

In the cloud, user probably does not know where his/her data is stored exactly. So controlling data storage is hard [26]. Most vendors do not know where the data is exactly stored, due to working with multinational companies. Moreover, sometimes accessing to these data for special processing such as recovery becomes impossible or hard. In globalization infrastructure, the country which date is stored is unknown, and users are concerned about national privacy regulations. For example, photos uploaded to Google+ can be stored in any servers around the world. For enterprise and private sectors it is important to know the location of storage, or even they prefer a location. This requires consent between vendors and users, but in most cases customers lack information about it [28]. It is possible that jurisdiction enforces to review data [29], e.g. UK's laws enforce providers to keep data within country [28].

IV. CONCLUSION

Emerging of cloud computing was forced by economic crisis and trends show it will be hot topic for following years. Some organizations are going to use this new model of computing without enough investigation. As a result of integrating IT in all strategic and operational aspects of organizations, failure of managing risks associated with cloud computing tragically will affect firm's survival.

Without having a comprehensive and detailed view of this model, deciding about it, can be completely cloudy and uncertain. Clarifying opportunities, which can be grabbed by gaining cloud computing offered benefits, will enlighten the choices.

In other hand, the security and risk are not new things in the computer age, but when we talk about cloud computing there are some addition cases that should be considered as risk. By identifying these risks, decision makers and risk management officers can deal with risk assessment in better understandable and less ambiguity environment.

The vendors should know of event that will happen in the future, accordingly, design a plan for such incidents. Nowadays, there are a lot of attack such as DOS and DDOS threatening the cloud computing, because so much data are stored in the cloud, and it is really worth trying to compromise this system. Security, privacy, integrity and availability are most important factors and users really concern about them.

As cloud, computing is still a new model; it needs more investigation and research to satisfy users concerns. We need to declare models of evaluating risks for using cloud computing and establishing frameworks, which can guaranty certain security levels.